# MOVING OBJECT TRAJECTORIES META-MODEL AND SPATIO-TEMPORAL QUERIES

Azedine Boulmakoul, Lamia Karim[1], Ahmed Lbath[2]

[1]FST Mohammedia, Département informatique
BP 146 Mohammedia 20650, Mohammedia, Morocco
azedine.boulmakoul@gmail.com
lkarim.lkarim@gmail.com
[2]Université Joseph Fourier, Grenoble
Laboratoire LIG BP 72- 38402 St Martin d'Hères, Grenoble, France
ahmed.Lbath@ujf-grenoble.fr

## ABSTRACT

*In this paper, a general moving object trajectories framework is put forward to allow independent applications processing trajectories data benefit from a high level of interoperability, information sharing as well as an efficient answer for a wide range of complex trajectory queries. Our proposed meta-model is based on ontology and event approach, incorporates existing presentations of trajectory and integrates new patterns like space–time path to describe activities in geographical space-time. We introduce recursive Region of Interest concepts and deal mobile objects trajectories with diverse spatio-temporal sampling protocols and different sensors available that traditional data model alone are incapable for this purpose.*

## 1. INTRODUCTION

Information and communication technologies advances have encouraged more and more mobility of moving objects (human beings, vehicles, etc.) with different heterogeneous sensors available (e.g. mobile phones equipped by GPS, vehicles with navigational equipments, cameras, Self Terminal), and multitude of GPS-enabled devices that are used in mobile phones, vehicles, bracelets, watches, laptops even in smart shoes. Today, the most interesting and attractive applications are location-based services, the demand of trajectories applications handling, visualizing and exploring moving objects trajectories is growing, e.g. a system for destination and future route prediction based on trajectory mining [1], real-time monitoring of water quality using temporal trajectory of live fish [2], analyzing bird migrations trajectory[3], searching for similar trajectories in spatial networks [4], building real-world trajectory warehouses [5], traffic mining [6], etc.

The number of raw data trace that is collected does increase exponentially so the problem of data accuracy and operating in an optimized way have become more complex. In this research, we propose a general moving object meta-model based ontology and event models to structure tremendous collected data. These traces are automatically collected and filtered for storage and operating in various fields.

By operating ontology, our proposed meta-model focuses on practical problems providing a shared understanding and common data model for different presentations of trajectories (raw, structured, semantic region of interest and space time path). Moreover, in our general meta-model, we use event approach to allow integration of information from heterogeneous spatio-temporal data sources with diverse spatial and temporal sampling protocols. Once our proposed moving objects trajectories meta-model has been instantiated, trajectories could be organized in spatio-temporal database, to support representing and querying of moving objects and their trajectories [7 and 8], or into trajectory data warehouse to analyze and make decision [5].

This research's goal is to provide a meta-model, for moving object's trajectories, based on space-time ontology and analytical geo-semantic, to allow several applications in trajectories domains benefit from information sharing as well as an efficient answer for a wide range of complex trajectory queries. Object approach, the Oriented Object Trajectory Meta model integrates previous models of geometric, structured and semantic of trajectory. Our model includes also the hybrid spatio-semantic model given in [9] and models the following, by using space-time event approach, analytical geo-semantic, and space time ontology:

- Spatial Model according to OGC Spatial Data Model.

- Observation domain of trajectory according to OGC Sensor Meta Model and OGC Feature Type.

- All activities between the beginning and the end of Space Time Path [10-12].

- Mechanism of detection used to collect generated positioning data.

- Movement patterns using composite Region of Interest.

The remainder of the paper is organized as follows: Section 2 presents basic concepts relating to trajectories of moving objects. We discuss, in section 3, the issues related to the representation of trajectories. In Section 4, we resume ontological models of events, space-time, semantic trajectory and space-time path paradigm. Section 5 focuses on presenting the proposed moving objects trajectories meta-model. Section 6 presents expressiveness query types for unified trajectory database. Finally, in Section 7 we conclude the paper and present some directions of future work.

## 2. BASIC CONCEPTS OF TRAJECTORIES

A trajectory is a description of physical movement of moving objects changing over time, in the following basic presentation of trajectories:

- Raw trajectory (figure 1) is the recording of the positions of an object at specific space-time domain, for a given moving object and a given time interval, it is presented as a sequence of geometric location in 2D spatial system ($x_i$, $y_i$, $t_i$).

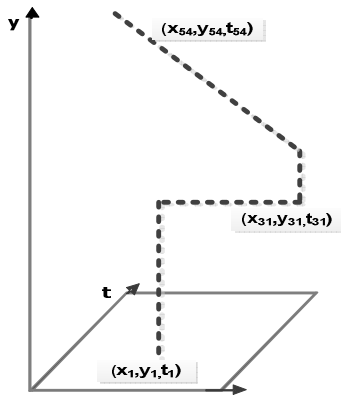

Figure 1.  Raw trajectory presentation

- Structured trajectory [3] (figure 2) defined as a raw trajectories structured into segments corresponding to meaningful steps in the trajectory trace (e.g. travel).

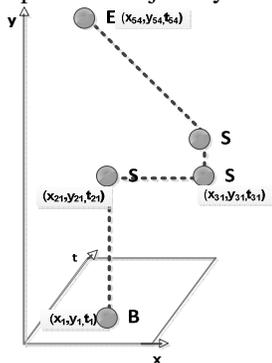

Figure 2.  Structured trajectory presentation

- Semantic trajectory [3] expresses the application oriented meaning using four component (stop, move, begin and end).  Stop, move, begin and end are no more spatio-temporal position, but semantic objects linked to general geographic knowledge and application geographic data  (figure 3).

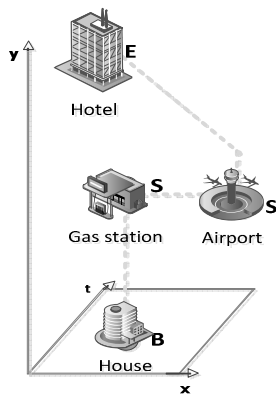

Figure 3.  Semantic trajectory presentation

Other recent approach describes movement patterns in both spatial and temporal contexts based on Region of Interest [13] by defining spatial neighbourhood and temporal tolerance (figure 4).

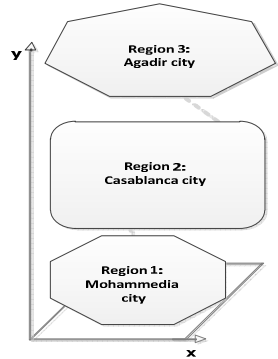

Figure 4.  Trajectory with Region of Interest

## 3. RELATED WORK

Meng and Ding [14] proposed a Discrete Spatio-Temporal Trajectory Moving Object Database system destined to model trajectory of moving objects by a set of straight lines within a constant speed. In [15], Wolfson presented a data model called Moving Objects Spatio-Temporal by representing the database location as a function of time and bound deviation. In [3], Spaccapietra stated that there are two facets of a trajectory: geometric facet which only considers the point geometry and a semantic facet which gives a meaning or semantic interpretation of application objects (Move, Stop, Begin and End). Studies in [16-17] focus respectively on the proper definition and analysis of semantic interpretations with stops and moves.

Patterns in [3] provide a trajectory structure as a sequence of moves and stops in between, with begin and end events to represent a trajectory in a relational model. However, there are many inconveniences when using a relational trajectory model, such as complex maintenance when upgrading application (a simple update of the application to a given point can have an impact in cascade with other functions of the application), and the application will then be edited in its entirety. Moreover, the functional approach is not suited to develop applications managing complex phenomena that are constantly changing (several tens of thousands of lines of code). Also, in view of enriching, the modelling of network-constrained trajectories, design pattern presented in [3] needs to further explore the interaction between trajectory modelling strategies and the multiple models that have been proposed in the literature. Furthermore, the current trajectory design pattern models trajectory just from the semantic point of view (as sequence of stops and moves).

Works proposed in [18] introduced a conceptual and computational approach for semantically trajectory data analyzing which redefines trajectory as the trace of a moving object that has geometric spatio-temporal and semantic features (the meaning of a movement). Study in [19] describes a framework for semantic trajectory relying on the definition of trajectory related ontology. Alvares [20] represented semantic trajectory via its link to application objects to better clean the initial data and to support more sophisticated methods that may significantly reduce the size of the trajectory dataset. Yan [9] proposed a hybrid spatio-semantic model and a computing platform for trajectories of moving objects.

Giannotti [13] describes movement patterns in both spatial and temporal contexts, based on static and dynamic RoI (Region of Interest) by allowing approximation in both spatial neighbourhood and temporal tolerance.

However, it is still very hard to explain and understand movement behaviours based on these patterns, because most of these models present trajectory and neglect having observation and description of trajectory/moving object (e.g. when taking a taxi or a bus, it seems to be very interesting to have information about full tank of gas in litters or how much gas it takes to fill up the tank). Also, working with geometrical facet is interesting for some users, e.g. some doctors need to have spatial coordinate (x,y) and time reference of his material for a chirurgical operation (lithotripsy, radiotherapy). For an appropriate terrorist track we need to find out in which region he could be. Whereas, marketing studies main interest for moving object's physical and virtual activities in a specific space-time. Furthermore, queries using current models database can not give detection's mechanism type used to collect data, e.g. which mechanism of detection was provided to collect data (space-time points, activities, etc.) of a person at 8 a.m?

Other benefit of the proposed data-model are introducing composite Region of Interest concepts and dealing Mobile Objects trajectories with diverse spatio-temporal sampling protocols and different sensors available that traditional data model alone are incapable for this purpose. Figure 5 presents an example of composite region of Interest, where region *Hypermarket* is composed of other regions: prepaid parking, bank, supermarket, department store and restaurant, respectively spatio-temporal data have been captured using different sensors GPS, RFID, data base transactions, and camera.

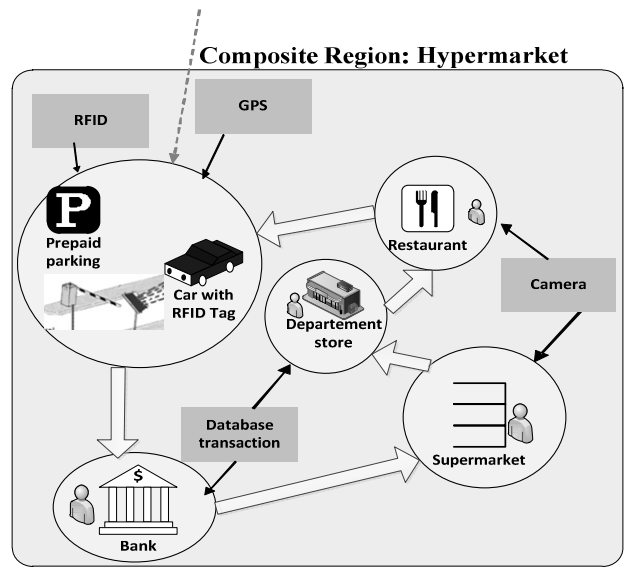

Figure 5. Trajectory with Composite Region of Interest and different sensors.

Below in table 1, we provide a brief tabular comparison of previous moving object's trajectories models used in literature with our proposed meta-model on the following criteria's basis: supporting presentation of Structured trajectory, Semantic trajectory, Region of Interest, Composite Region of Interest, Activities/ space time path, Model type, Mechanism of detection provided, Based ontology and Event approach.

Table 1. Comparison of previous models of moving objects trajectories.

| | Structured trajectory | Semantic trajectory | Region of Interest | Composite Region of Interest | Activities/ space time path | Model type | Based ontology and Event approach | Mechanism of detection provided (other than GPS) | OGC Sensor meta-model | OGC Observation feature-type |
|---|---|---|---|---|---|---|---|---|---|---|
| DSTTMOD [14] | Yes | No | No | No | No | Relational | No | No | No | No |
| Moving Objects Spatio-Temporal [15] | No | No | No | No | No | Function of time and bound deviation | No | No | No | No |
| Conceptual view on trajectories [3] | Yes | Yes | Yes | No | No | Relational | No | No | No | No |
| Hybrid spatio-semantic model [9] | Yes | Yes | Yes | No | No | Relational | No | No | No | No |
| Region of Interest [13] | Yes | No | Yes | No | No | T-Patterns mining | No | No | No | No |
| Our proposed meta-model | Yes | Yes | Yes | Yes | Yes | Object oriented | Yes | Yes | Yes | Yes |

## 4. SPACE-TIME ONTOLOGY

In our proposed trajectories meta-model we use ontology of events to give new simple and usable way for trajectories data modelling and visual analytics. In the following, we summarize events description, sources and ontological model.

### 4.1. Events' ontology

Event ontology has already been proven useful in a wide range of contexts, due to its simplicity and usability. Several ontology describing "events" and related concepts have been investigated. The SHOE General Ontology [21] defines an event as something that happens at a given place and time. In Dublin Core metadata standard [22], an event is defined as 'a non-persistent, time-based occurrence'. Quine [23] described events as objects where objects are regions bounded in space and time.

Principal components of event's ontology are: (a) Event to describe "what" is happening. (b) Actor to describe "who" is doing the action. (c) Time to describe "when" is something happening: to range precise dates, estimated Time of Arrival or imaginary time spans. (d) Place to describe "where' is something (occurrence or happening) happening (geographical coordinates), Event Ontology recommends the use of GeoNames to identify a place: anonymous places, identified with latitude and longitude. (e) Object to describe "with what" is something being done.

## 4.2. Sources of events

Today's technological advances of sensors have enabled capturing events easily and with better precisions. Position, temperature, speed, force, and pressure sensors collect information on the behaviour of the operative part and turn it into information usable by the control part. Information is an abstract quantity that specifies a particular event among a set of possible events. In order to be processed, this information will be carried by a physical medium (energy) called a signal.

## 4.3. Space, time and theme ontology models

Study in [24] has provided upper-level ontology for modelling three dimensions of data: thematic, spatial, and temporal. They combined concepts and relationships from both the thematic and spatial dimensions, and show how to deal temporal semantics into there ontology. To represent two-dimensional space and time information in the model, they [24] distinguished between entities which persist over time and maintain their identity through change named continuants, and events that happen and then no longer take place called occurrents. Spatial dimension presented by Spatial Region, Coordinate, and Coordinate System entities. Temporal dimension presented by associating time intervals with relationship instances.

The main concepts presented in Matthew's ontology model [25] are: (a) Geographical Place to represent a geographic feature or a named place. (b) Named place linked to Footprint spatial element to geo reference point, line, or polygon. (c) Dynamic Entities represent undefined spatial properties. (d) Entities with static spatial properties like buildings, administrations, markets, universities or a city are presented by Named Places entities. (e) Events are special types of entities which represent occurrences in space and time e.g. Workshop, inauguration of an institute, etc.

Event based models and approaches that have appeared in the literature presented a categorisation of change. Kate Beard in [26] describes an event based approach to model space and time, instead of presenting it using geographic and identities characteristics which could be repeated, where dynamic aspect dominates instead of geographical features. In this approach, change or what we call event is the core of space time model. This model presents an event as a localisation in space and time. A space time setting has the subclasses spatial, temporal and spatial-temporal. The attributes of a primitive event depend on the change type (appearance, transformations, movement).

## 4.4. Modelling Space and Time in Location-Based Services

Space time location and movement of a moving object are basic characteristics in all Location-Based Services (LBS) applications. In [27], Dieter proposed an LBS ontology structure composed of (i) Domain Ontology using the Space Ontology and Time Ontology, (ii) the Content Ontology, focusing on the LBS specific content e.g. restaurant, conference hall, park, and (iii) the Application Ontology including the profile ontology for the LBS, e.g. tourist, sport fan, businessman, family, and service ontology e.g. route to restaurant, closest restaurant. Basic classes used in [27] to present domain ontology Knowledge-Base of LBS are:

- Space Ontology, constituted of classes of Location (position and reach of the moving object), Position (absolute (x,y) coordinates), Point, Reach (the surrounding area of position), Uncertainty (coordinate deviation for the accurate value) and Movement.
- Time Ontology, composed of classes of Time (when an action or event happens), Timestamp, Uncertainty (time deviation for the accurate value) and Movement.
- Content class, which is further related to the Content Ontology.

## 4.5. Semantic trajectory ontology

Recent researches in modelling trajectories are interested to study and analyse not only time geographical location (coordinates and regions) but also semantic side of trajectories. Presenting semantic trajectories, by adding non geometric knowledge to moving objects trajectories data, give more semantic, interpretation and meaning to movement. Semantic facets of trajectory explicitly expressed by adding a semantic layer linked to application need [3].

Semantic trajectory ontology for a traffic management application [19] consists of three ontological modules: the geometric trajectory module, the geography module, and the application-domain module.

- The geometric trajectory module describes spatio-temporal features, as a set of structured trajectories: *Begin*, *End* and sequence of *Stops* and *Moves are* structures components.
- The geography module is linked to geometric trajectory and application modules, because each geometric element has a geographical corresponding according to application domain. This module constituted from building and places, topography and network relations.
- Application domain module describes all concepts related to application.

## 4.6. Space-time path paradigm

Information and communications technology advances have brought important changes to the way activities are carried out. For example, GeoLife, a location-based social networking service developed by Microsoft Research Asia [28], facebook, twiter, enable people to share their activities. Geolife is a transportation mode detection system, which classifies a segment of a GPS trajectory into one of {"walking", "driving", "biking" and "onBus"}.

Adding activities to trajectory presentation, like transportation mode, allow users to have an extended query type, like asking how a person X went to work? Or how much time this person spends in walking every week?

Activities studies have developed systems for knowledge discovering to understand people behaviours, and other phenomena. Many researches have been interested in activity recognition from trajectory data in [29]. Many spatial and temporal studies have used the Hägerstrand's framework to study characteristics of human activities in physical space. Shaw [30] extended concept of space-time path, shown in figure 6, to represent both physical (walking, driving, etc.) and virtual activities (sending email, calling through a mobile phone). As each activity has a geographical location and time interval, space-time path has been profiled as a container of all activities occurring by a moving object.

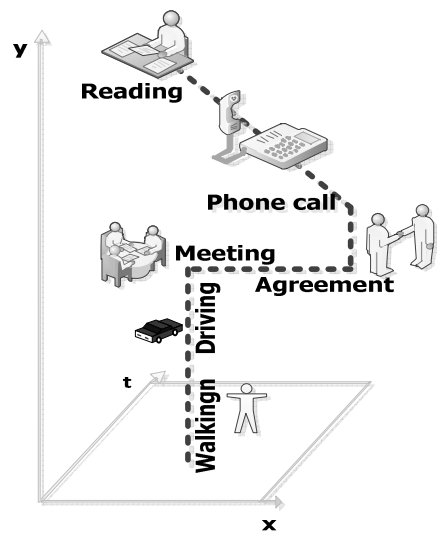

Figure 6.  Space-time path presentation.

## 5. THE PROPOSED UNIFIED TRAJECTORIES META-MODEL OVERVIEW

In our proposed meta-model of trajectories, we use object approach to increase developer's productivity and deliver high-quality applications: modularity, reusability, genericity and extensibility. Instantiation of our Meta model allows producing existing trajectories models as well as allowing creation of space-time path concept (trajectories sequence's of activities). According to our meta-model, regions of interest are composites though it may describe modal transportation networks, Voronoi spatial networks, etc. In the following we will give some modelling elements as packages in order to discuss underlying global idea.

### 5.1. Package diagram for our trajectory patterns

In this section, we use UML package diagram to group elements and make UML diagrams simpler and easier to understand. The most important packages used in our trajectory patterns are described as follow.

#### 5.1.1. Space Time Path Domain package

Space Time Path Domain, the central package, combines classes of Raw, Structured, Semantic Trajectory and Space Time Path to provide the different models of trajectory that can be used.

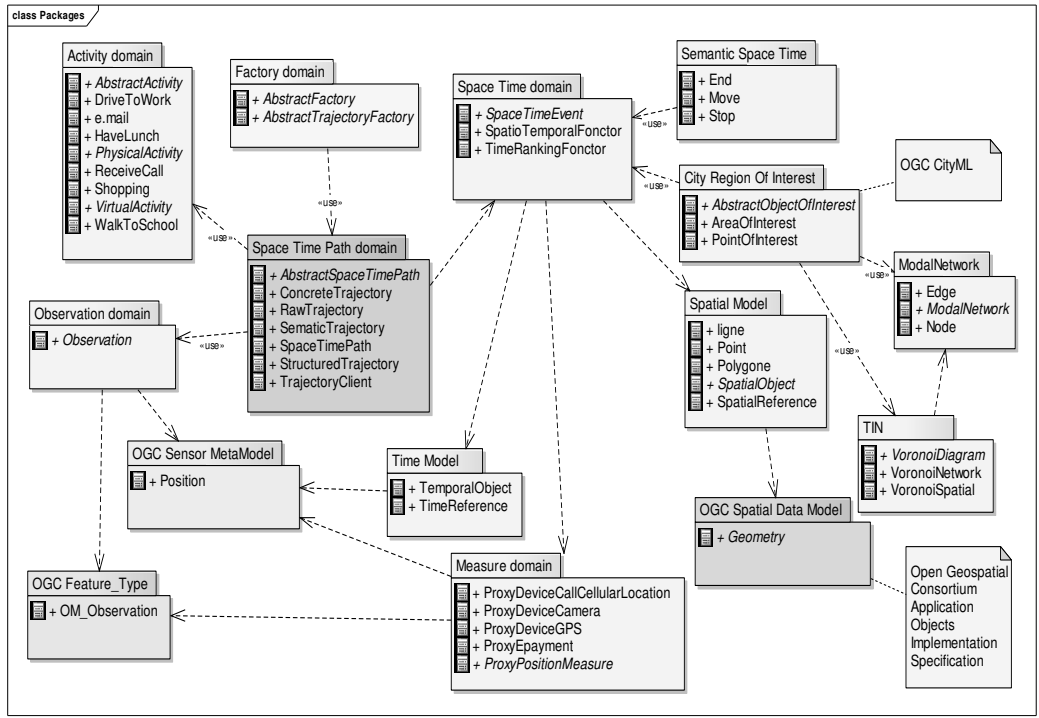

Figure 7.  Package diagram for proposed trajectories patterns

### 5.1.2. Activity Domain package

As monitoring activity of moving objects is a crucial task in many urban, economic and social systems applications, we added a UML package Activity Domain including classes of physical activities (e.g. drive to work, have lunch, drive to school) and virtual activities (e.g. send email, receive a call). This package allows us to represent, manage and analyze activities that occur in physical and virtual spaces, events that involve multiple individual's activities and projects that consist of multiple events, also objects of this package are used to create space-time path [10-12].

### 5.1.3. Observation Domain package

In the context of Open Geospatial Consortium Sensor (OGC) meta-model and OGC Feature type, we added, to our trajectory meta-model, Observation Domain package to describe action with a result describing some phenomenon. For example, when tracking a terrorist's trajectory, it may be important to record observation of activity on the Internet, the amount of email sent to a terrorist forum, how terrorist set up free e-mail accounts, write messages and save them in draft form, so that other users with access to the same account can read the email without ever having to send it (Emails are only interceptable when in transit).

### 5.1.4. Measure Domain package

GPS equipped mobile phones, device camera, vehicles with navigational equipment or location based services, give digital traces (sequences: $<(x_1,y_1,t_1)$, ..., $(x_n,y_n,t_n)>$) to describe mobility behaviour (figure 1). In order to define the basic models for how geospatial information is to be characterized and encoded according to the OGC, these devices are arranged in Measure domain package which use OGC Sensor MetaModel package and OGC Feature_Type package.

### 5.1.5. Region of Interest packages

In many applications, it is expensive to work with trajectory as a sequences of Spatio-temporal events: $<(x_1,y_1,t_1), ..., (x_n,y_n,t_n)>$. Giannotti [13] describes movement patterns in both spatial and temporal contexts, based on static and dynamic Region of Interest (RoI), by allowing approximation in both spatial neighbourhood and temporal tolerance.

In this package, we model RoI and introduce a composite RoI class, e.g. in a tourist guide application, we can consider Morocco as a RoI, however in other user's request; the RoI could be the city of Casablanca, monument or beach. Hence in our meta-model, a RoI could be composite of one or several RoI. Also, we modelled RoI as a Voronoi polygon in which a site can represent a point of interest.

## 5.2. Class diagram for our trajectory meta-model

To provide an overview of our proposed meta-model of trajectories, we present in this part class diagram for trajectories. In the following, we present needed classes to be instantiated for modelling and producing different models integrated and modelled in our proposed meta-model.

### 5.2.1. Producing Structured Trajectory Model

For a given moving object and time interval, GPS allows collecting tremendous amounts of spatio-temporal coordinates $(x_i, y_i, t_i)$. Modelling trajectories with structured presentation accelerated processing of application that only referring to the information about the spatio temporal position of the beginning B, end E and sequence of stops S and moves M are stored. Figure 2 illustrates the structured presentation of trajectory.

Class diagram, shown on figure 8, presents classes to instantiate for modelling a structured trajectory model. Below is a description of each class of this instance:

SpaceTimeEvent class presents an event as an occurrence that happens in a small space and lasts a short time, e.g. drinking water, changing position, connect answer machine. From temporal point of view, SpaceTimeEvent class is a composition of TemporalObject class according to specification of TimeReference class. From spatial point of view, it is a composition of SpatialObject according to SpatialReference class specification.

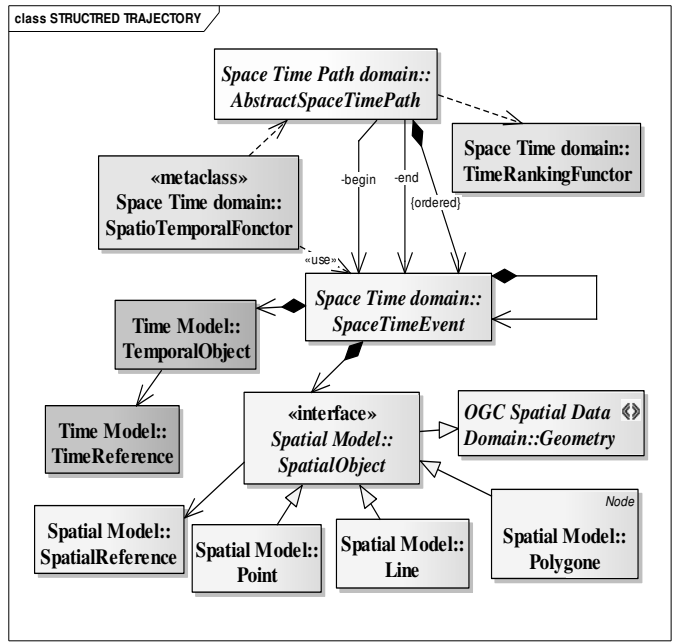

Figure 8.  Structured Trajectory Model class diagram

Each SpatialObject class is a specialisation of OGC Spatial Data Domain::Geometry class. In our model, we present SpatialObject as polymorphic class, to make derived classes code easier to change (point, line or polygon class). Using UML notation, we model reflexive composition relation where the parent and child are basically from the same SpaceTimeEvent class. For example, workshop could be considered as an event but inside this event we find other events, like Registration/Breakfast, Invited Talks, Coffee Break, End of Workshop, etc.

AbstractSpaceTimePath class traces the path taken by a moving object, it is a composition of ordered SpaceTimeEvent class, have begin SpaceTimeEvent and end SpaceTimeEvent.

TimeRankingFunctor class added to order times whereas SpatioTemporalFonctor class uses SpaceTimeEvent class to order SpaceTimePaths.

### 5.2.2. Producing Semantic Trajectory Model

Although using structured presentation is better than raw presentation in terms of performance and querying speed, this presentation is poor of meaning. Analyzing this trajectory is hard and cannot answer complex query like: find cars license plate when they are usually park at the parking lot.

It is interesting to add more semantic information to give a meaning to trajectory, e.g. person at 7 a.m. was in his/her house, and then headed to school at 8 a.m. She made it to school, at 9 a.m. then got to the bus station at 10 a.m. and at the end of her trajectory she stopped in the stadium at 11 a.m. (figure 3).

By definition, a semantic trajectory is a structured trajectory with added semantic information, figure 9 shows semantic trajectory class diagram. In addition to structured trajectory classes (AbstractSpaceTimePath, SpaceTimeEvent, SpatialObject, TimeRankingFunctor …), user needs to add SemanticSpaceTimeEvent class, which is a specification of SpaceTimeEvent, each SemanticSpaceTimeEvent is SpaceTimeEvent with semantic information that have begin and end

SpaceTimeEvent, Begin, End, Stop and Move are no more spatio-temporal positions, but SemanticSpaceTimeEvent where moving objects begin, end, stop and move.

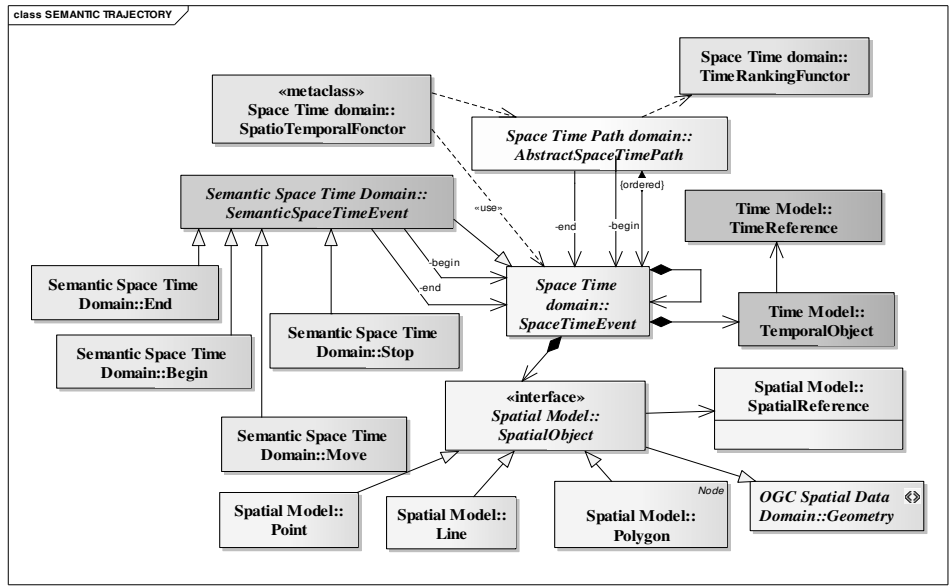

Figure 9. Semantic Trajectory Model class diagram

### 5.2.3. Producing Trajectory with Composite Region of Interest model

As shown in the second section, Trajectory with Region of Interest model describes movement patterns in both spatial and temporal contexts based on Region of Interest. For example, a person trajectory could be presented as follows: Person X was at 9 a.m. in neighbourhood "R1" then moved to neighbourhood "R2", so he finally made it to his last trajectory which was the neighbourhood "R3" at 10 a.m. Our meta-model allows creating this model of trajectory, also we modelled that a region of interest could be composed of other regions of interest. For example, to get details and precisions, a region of interest like supermarket should be composed into others RoIs: food region and non food region. Also, in a non food region we find other regions like cosmetics, flowers, books, clothes and toys departments (regions). Class diagram, in figure 10, presents instantiated trajectory with Region of Interest class diagram.

A set of SemanticSpaceTimeEvent class are located in AbstractObjectOfInterest class. Using the inheritance relationship, AbstractObjectOfInterest could describe a PointOfInterest (supermarket, bank…), area of interest (industrial region, tourism region), modal network, and Voronoi diagram. We added reflexive composition Relationship on the AbstractObjectOfInterest class to model that an ObjectOfInterest is composed of other ObjectOfInterest.

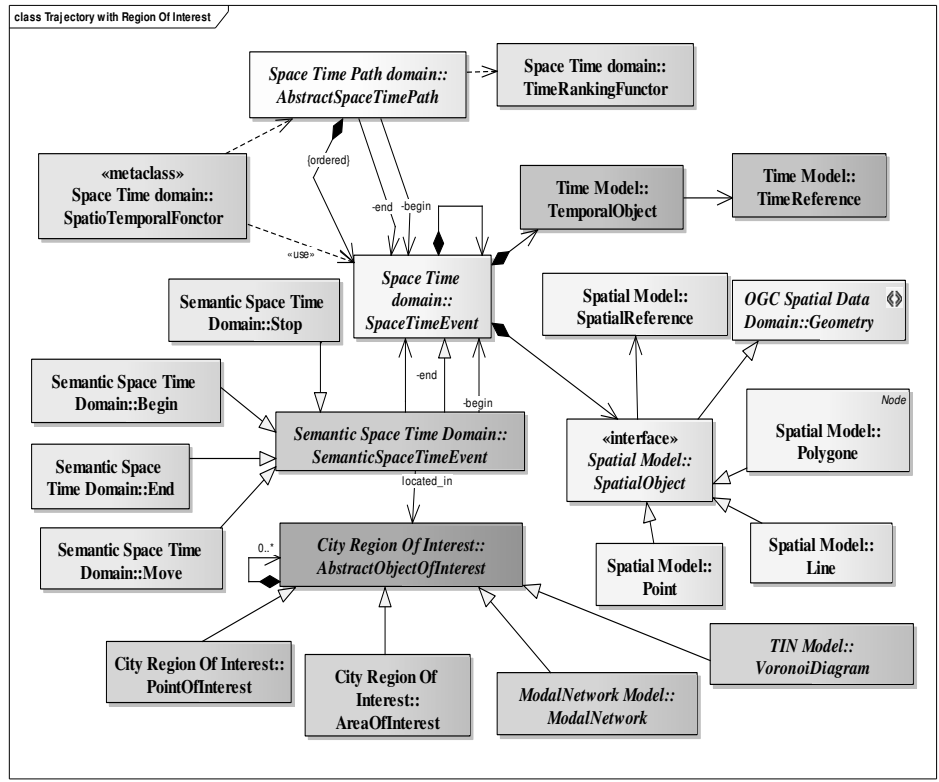

Figure 10. Trajectory with composed Region of Interest class diagram

### 5.2.4. Producing Space Time Path Model

Space Time Path is defined by the sequence of both virtual and physical activities describing a moving object in an integrated space-time. In our proposed meta-model, we extend raw, structured, semantic models of trajectory and trajectory based on region of interest to integrate activities and processes of a moving object. Figure 6 gives an example of space time path of a moving object.

Space Time Path class diagram, shown in figure 11, allows creating a Space Time Path from previous models of trajectories raw, structure, semantic trajectories and trajectory with region of interest.

To get visibility of activities of moving object, at each space time, we have added between these two following classes SpaceTimeEvent and AbstractActivity a beginning and ending activity associations. Thanks to UML Generalization relationship, we have modelled both virtual activity, like sending email and receiving a call, and physical activity like drive to work, walk to school or shopping. Also, in order to support spatio-temporal analysis, we added a composition relationship between process and activities to capture activities as a process.

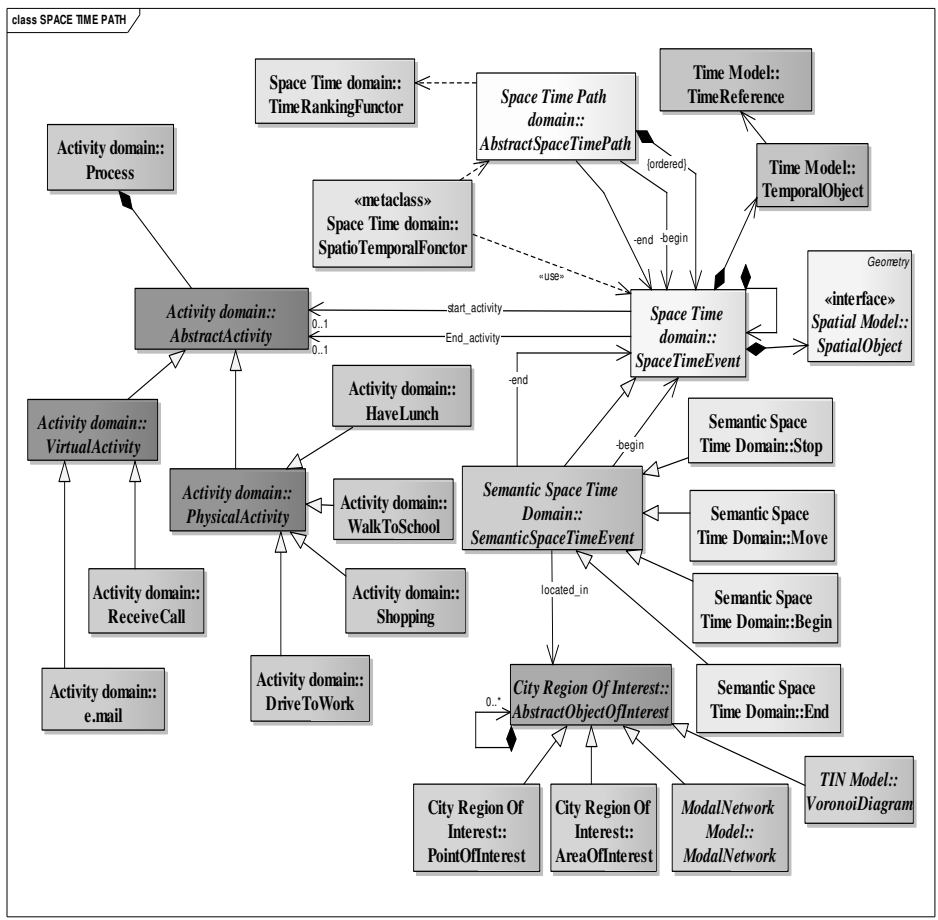

Figure 11. Space Time Path Model class diagram

### 5.2.5. Producing Space Time Path Model Trajectory Observation Measure Domain Model

In figure 12, we present our trajectory Observation Model to store pertinent and useful observations. To get observations of a moving object at each space time, we created Observation model; it consists of association between SpaceTimeEvent class and Observation class according to OGC feature Observation reference class.

Trajectory Measure Domain Model have been integrated not only to get information about mechanism of detection, used to collect generated position data, at a specific space time, but also to make it possible to analyze the reliability of data collected.

ProxyPositionMeasure class, presented in figure 7, functioning as interface to several measure devices: ProxyDeviceCamera, ProxyDeviceCallCellularLocation, ProxyDeviceGPS and ProxyEpayment. Thanks to the association between a SpaceTimeEvent class and ProxyPositionMeasure class, we can know, at any moment, device's characteristics used to capture data, whatever the model used to present the trajectory.

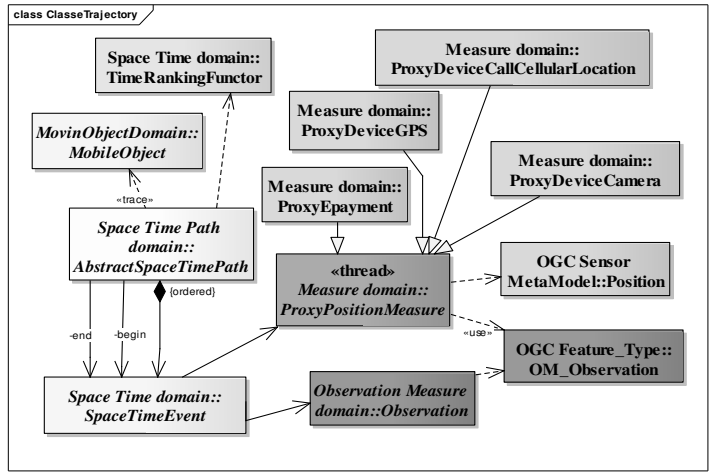

Figure 12.  Trajectory's observation and measure domain class diagram

## 5.2.6. Trajectory's Factory Model

The goal of the proposed abstract trajectory factory pattern, shown in figure 13, facilitates the use of our pattern and guides our model users. In the following, we highlight this factory classes: AbstractTrajectoryFactory class encapsulates different presentations of trajectory (Raw Trajectory, Structured Trajectory, Semantic Trajectory, Trajectory with Region of Interest and Space Time Path), whereas the ConcretTrajectory class is the location which creates and constructs trajectories presentations. The intention in using AbstractTrajectoryFactory and ConcretTrajectory classes is to make our pattern extensible; i.e. we can add new presentation of trajectory with no change to the code uses the base class, user codes interfaces only with TrajectoryClient class and AbstractTrajectoryFactory to access to a trajectory object.

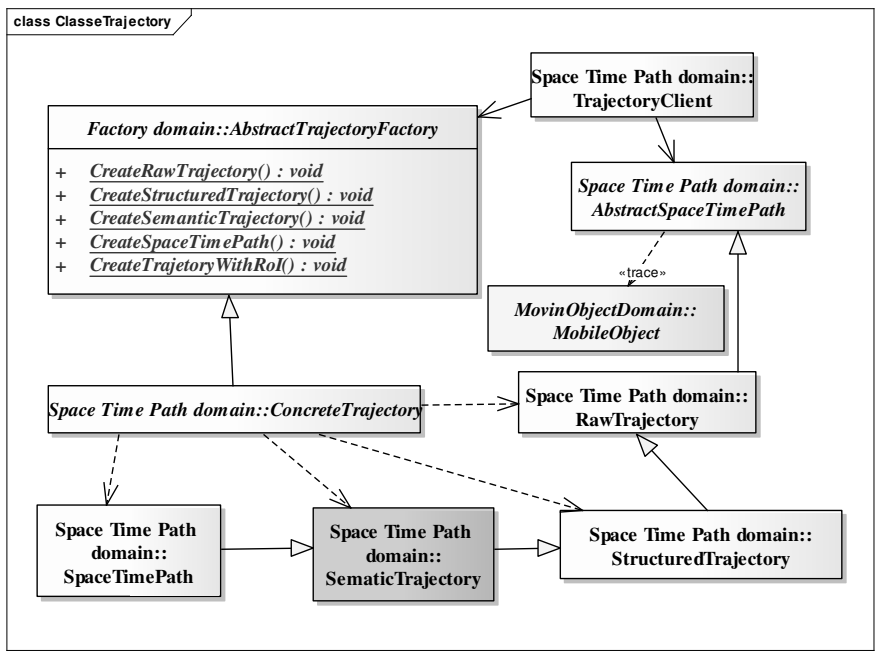

Figure 13.  Trajectory's factory class diagram

# 6. UNIFIED MOVING OBJECT TRAJECTORY QUERIES

Our work differs in the fact that we provide a unified model for trajectories data exploiting to answer a wide range of complex trajectory queries. For example, query like "cars license plate that park in front of my house noticing it all, when I am on the way to school" can not be answered using traditional models.

In this paragraph, we evaluate efficiency, performance and utility of models instantiated through a set of spatio-temporal queries. Using our proposed unified meta-model, trajectory queries can be classified into six types according to their instantiated spatiotemporal data-model:

- *Raw trajectory queries*: ask for spatio-temporal coordinates of a specified moving object (MO) at a given time t to specified trajectory segment(s). Asking for semantic information need using of spatial join. e.g. find all places (restaurant, supermarket and administrations) visited by a moving object (MO):

*select r.name*

*from rawtrajectory t , restaurant r*

*where t.id='MO' and intersects (t.spatialpoint.geometry , r.geometry)*

*Union*

*select s.name*

*from rawtrajectory t , supermarket s*

*where t.id='MO' and intersects (t.spatialpoint.geometry , s.geometry)*

*Union*

*select a.name*

*from rawtrajectory t , administrations a*

*where t.id='MO' and intersects (t.spatialpoint.geometry , a.geometry)*

- *Structured trajectory query:* asks for spatio-temporal coordinates where the moving object's stop, move, beginning and end, in specified trajectory segment(s). E.g. finding all roads with moving object (MO) when it took him/her over 10min.

*select r.name*

*from StructuredStop t , road r*

*where t.id='MO' and intersects (t.spatialpoint.geometry , r.geometry) and (t.timeEndStop - t.timeBeginStop )>10*

- *Semantic trajectory query*: ask for trajectories where moving object (MO) stayed in a given semantic place (restaurant, cinema, stadium…) for a while (e.g., 1 hour). E.g. find all trajectories where moving object MO got on the road and took him/her over 10min.

*select t.name*

*from SemanticStop t*

*where t.id='MO' and t.cat='road' and (t.timeEndStop - t.timeBeginStop)>10*

- Trajectories and regions of interests query: ask for trajectories crossed a point of interest, area of interest, modal network or voronoi diagram in a given time interval, e.g. Figuring out the number of trajectories that visited each commercial region.

*select t.nameRegion, count(*) nb_visits*

*from RoITrajectory t*

*where t.cat='commercial'*

*group by t. nameRegion*

- *Space time path query*: ask for activities or process of a moving object in a spatio temporal location, e.g. finding the space time path of a specific person.

*select t.name, t.time, t.physicalActivity, t.virtualActivity*

*from SpaceTimePath t,*

*where t.id='MO'*

- *Trajectories and mechanism of detection query:* ask for devices and their reliability degree used to capture information in a specific spatio-temporal location, e.g. which mechanism of detection used to capture information when the moving object was at the airport.

*select d.name, d.reliability*

*from RoITrajectory t , Devices d*

*where t.name like='%airport%' and t.id= d.tid and t.time=d.time*

## 7. CONCLUSIONS

Based on space time ontology and events approach, we proposed a generic meta-model for trajectories of moving objects. The present work has a double aim. First, providing a higher level of interoperability and information sharing, where we happen to have traditional trajectories models which they are, raw, structured and semantic as they have been integrated using ontology approach.

The second aim is to extend previous data models, for knowledge discovering and clearly answering a wide range of complex queries, as it shows us new interesting patterns (i) space–time path to describe physical and virtual activities of a moving object, (ii) recursive region of interest, etc.

In future work, we are going to provide a suitable system to implement our meta-model and project it into a useful application domain.